# Direct Numerical Simulation of high Prandtl number fluids and supercritical carbon dioxide canonical flows using the spectral element method

**Tri Nguyen, Elia Merzari**
Ken and Mary Alice Lindquist Department of Nuclear Engineering
The Pennsylvania State University, 228 Hallowell, University Park, PA 16802, USA
nguyen.tri@psu.edu, ebm5351@psu.edu

**Haomin Yuan**
Nuclear Science & Engineering Division
Argonne National Laboratory, Building 208/C145, 9700 S Cass Ave, Lemont, IL 60439, USA
hyuan@anl.gov

**ABSTRACT**

The design of advanced nuclear reactors (Gen IV) involves an array of challenging fluid-flow issues that affect safety and performance. Currently, these problems are addressed in an ad-hoc manner at varying scales which are time-consuming and expensive. The creation of a high-resolution heat transfer numerical database has the potential to help develop to accurate and inexpensively reduced resolution heat transfer models. Such models can help address industrial-driven issues associated with the heat transfer behavior of advanced reactors. The models can be developed using the multiscale hierarchy developed as part of the recently DOE-funded center of excellence for thermal-fluids applications in nuclear energy. Ultimately this can lead to fast-running reliable models, thus accelerating the deployment of advanced reactors. In this paper, we performed a series of Direct Numerical Simulation using the spectral element codes Nek5000 and NekRS to investigate heat transfer in mixed convection conditions. First, we investigate the heat transfer of the flow in heated parallel plates for high Prandtl number fluids. The calculated database will eventually be used to evaluate existing heat transfer correlations and some modifications will be proposed for cases where no satisfactory choice is available. We have also investigated the heated transfer alteration phenomena in a straight heated tube for supercritical carbon dioxide ($sCO_2$). The low-Mach-number approximation is used to decouple thermal and dynamic pressure, as pressure drop is negligible in this problem. The properties of $sCO_2$ are calculated using multi-region polynomials. We observed that the heat transfer deterioration occurred in combination with the property changes of $sCO_2$ and the depreciation of turbulence kinetic energy (TKE) for upward flow. Whereas, in downward flow, the heat transfer is enhanced thanks to the increase of TKE.

## 1. Introduction

Turbulence modeling historically has been a very challenging problem even for traditional coolants like water. Modern reactor designs often use non-traditional coolants, which has heat transfer characteristics where traditional water-based correlations (e.g., Prandtl unity heat transfer correlations) cannot be applied. Currently, there is a huge gap of DNS data that can be used to advance reactor design. The experimental results and high-fidelity numerical simulations which could be used to bridge the gap are ineffectively implemented. Recently, Center of Excellence for Thermal Fluids Applications in Nuclear Energy [1] has been established as a collaboration between Idaho National Laboratory and Argonne National Laboratory. A component of COE as a university consortium led by Penn State University has been established to develop and validate reliable and affordable advanced thermal-hydraulic models. The consortium has identified turbulence modeling for heat transfer as an industrial driven challenging problem. The DNS data generated will be used to develop reduced resolution models and evaluate existing heat transfer correlations.

In this work, we will perform Direct Numerical Simulation (DNS) for forced and mixed convection heat transfer using the Spectral Element codes Nek5000 [2] and NekRS [3]. NekRS is a novel refactor of Nek5000 [2] oriented toward the use in GPU architectures. The downcomer can be presented as heated parallel plates. The first test case is a simplified geometry representative of the downcomer of advanced reactor designs. The second test case is the flow in pipes using supercritical $CO_2$ ($sCO_2$) as the coolant.

For the first test case, a wide range of Ri ranging from 0 (forced convection) to 400 has been investigated. The working fluid is FLiBe, a high Pr molten salt mixture of lithium fluoride (LiF) and beryllium fluoride ($BeF_2$) [4] which is the interest to our industrial partner Kairos Power. Forced convection cases data has been compared with available DNS data of Kasagi et.al [5,6] and good agreements has been achieved. It is important to note that for high Pr fluid, the momentum diffusivity dominates the thermal diffusivity, and the momentum boundary layer is thicker than the thermal boundary layer. The Bachelor length scale must be used to estimate the resolution requirements, it can characterize the diffusion of heat by molecular processes which cannot be achieved by using Kolmogorov length scale. The Bachelor length scale is the ratio of Kolmogorov length scale to the square root of Schmidt number:

$$\eta_T = \frac{\eta}{Sc^{\frac{1}{2}}}$$

For heat transfer, the Pr number can be used instead of the Schmidt number:

$$\eta_T = \frac{\eta}{Pr^{\frac{1}{2}}} = \left(\frac{\alpha^2 \nu}{\epsilon}\right)^{\frac{1}{4}}$$

Unlike the use of Kolmogorov length scale for low Pr fluids DNS [7], the main drawback of using Bachelor length scale is that the spatial resolution for high Pr fluid like FLiBe can be orders of magnitude higher than for low Pr fluids, which can result in drastically increase of the computational cost. Traditionally, Reynolds Averaged Navier Stokes (RANS) turbulence models are applied to overcome the computational cost of DNS. However, the use of RANS turbulence models can propose significant uncertainty due to the restrictive and limiting assumptions of their formulation. In particular, most RANS models were developed based upon shear-generated turbulence mechanisms and they are all suited for mixed convection applications. However, additional terms must be introduced to account for buoyancy-produced turbulence, and these corrections require adequate data for testing and validation [8,9]. Furthermore, most of the RANS models of common application rely on the Boussinesq eddy viscosity concept [10] and employ a linear relationship between the anisotropic stresses and the velocity strain rate tensor. Because of the intrinsic limitations in RANS models, advanced turbulence models are often implemented and compared with high-fidelity validation data sets, which are typically provided by DNS data. The generated DNS data can assess RANS model uncertainty in cases involving the interaction of multiple physical phenomena.

Beside the heated parallel plates, we also investigated as a second test case, the heat transfer alteration phenomena in a straight heated tube for supercritical carbon dioxide ($sCO_2$). $sCO_2$ has been attracting attention in recent decades due to the growing interest of using $sCO_2$ as working fluid in advanced power conversion cycles. Due to the property changes in $sCO_2$, many studies have been trying to find empirical correlations between supercritical fluid flow and heat transfer deterioration. However, a quantitative prediction of the heat transfer deterioration is still challenging, and thus more fundamental understanding of supercritical fluid flow and heat transfer deterioration is still needed and can only be achieved by DNS.

For sCO$_2$ study, NekRS has been used to perform DNS for sCO$_2$ flow through a straight heated tube under mixed convection conditions. Heat transfer characteristics is evaluated by implementing low-Mach-number approximation. The thermal and dynamic pressure is decoupled, as pressure drop is negligible in this problem. We validated our wall temperature calculation result with the pioneering paper by Bae [11] and a good agreement has been achieved.

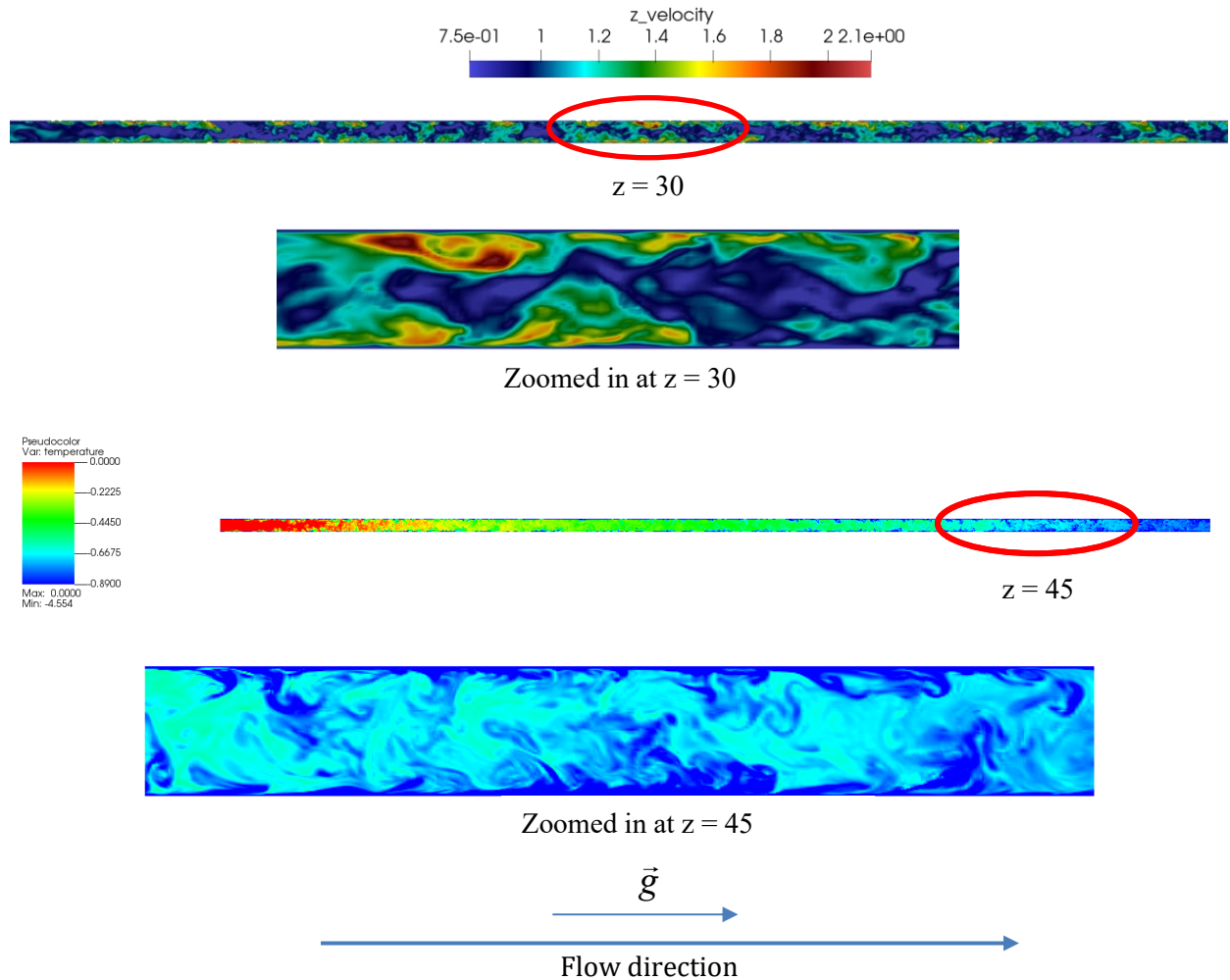

**Figure 1.** An example of velocity and temperature fields of symmetric cooling parallel plates with mix convection.

## 2. Numerical Methodology

### 2.1. Numerical Method

Nek5000 and NekRS, the primary tools used in this work, are ideally suited for large eddy simulation (LES) and DNS of turbulence. A new version of Nek5000 called NekRS has been recently developed for GPUs [3]. NekRS can utilize effectively GPU architectures and thus drastically improve the calculation speed. Over the last decades, Nek5000 has been receiving extensive validation in numerous references [12,13,14,15], demonstrating its capability to perform high fidelity LES and DNS in both simple and complex geometries. In this part for heated parallel plates, we performed a series of DNS with the consideration of the incompressible Navier-Stokes (NS) equations of a Newtonian fluid subject to buoyancy, modeled though the Boussinesq approximation:

$$\nabla \cdot u = 0 \tag{1}$$

$$\rho\left(\frac{\partial u}{\partial t} + u \cdot \nabla u\right) = -\nabla P + \mu \nabla \cdot \nabla u + \rho g(1 - \beta(T - T_0)) \tag{2}$$

$$\rho c_p \left(\frac{\partial T}{\partial t} + u \cdot \nabla T\right) = \nabla \cdot (k \nabla T) \tag{3}$$

where Eq. (1) is mass continuity, Eq. (2) represents momentum conservation and Eq. (3) is the energy conservation equation expressed in terms of temperature. u and g are vector velocity and gravity, respectively, $\rho$ is the density of the fluid, $\mu$ represents the dynamic viscosity, $\beta$ represents the heat expansion coefficient, k is the conductivity and $c_p$ the heat capacity. The Navier-Stokes (NS) equations are solved using the spectral element method (SEM). The SEM, introduced by Patera [16] is a subclass of the Galerkin methods, or weighted residual methods. It is a high-order method that provides good geometric versatility comparable to finite element methods, but also with minimum numerical dispersion/dissipation and rapid convergence properties typical of spectral methods [2]. Nek5000 is highly scalable [17,18,19] and has been using in many works to gain new insight into the physics of turbulence in complex flows.

In SEM, the domain is discretized into E hexahedral elements and represents the solution as a tensor-product of $N^{th}$-order Lagrange polynomials based on Gauss-Lobatto-Legendre (GLL) nodal points, which results in $E(N+1)^3$ degrees of freedom per scalar field. GLL points are chosen for efficient quadrature for both velocity and pressure spaces. The pressure can be solved at the same polynomial order of the velocity $N$ ($P_N$ -$P_N$ formulation) or at lower order $N$-$2$ ($P_N$ – $P_{N-2}$ formulation). Two time-stepping schemes, both up to third order, are available: BDF and OIFS [20]. The latter has the advantage of less severe stability limitation allowing for CFL > 1, but it results in a larger cost per time step.

In this work, the nondimensional form of N-S equations (1), (2), (3) are applied:

$$\nabla \cdot u^* = 0 \tag{4}$$

$$\left(\frac{\partial u^*}{\partial t^*} + u^* \cdot \nabla u^*\right) = -\nabla P^* + \nabla \cdot \frac{1}{Re} \nabla u^* - Ri * T^* \tag{5}$$

$$\rho^* \left(\frac{\partial T^*}{\partial t^*} + u^* \cdot \nabla T^*\right) = \nabla \cdot \frac{1}{Pe} \nabla T^* \tag{6}$$

Where $u^*$, $t^*$, $\rho^*$, $P^*$, $T^*$, Re, Ri, Pe are nondimensional velocity, time, density, pressure, temperature, Reynolds number, Richardson number and Peclet number, respectively. $u^*$=u/U, U is the inlet bulk velocity, $t^*$ = t/(D/U), D = 4$\delta$, D is the hydraulic diameter, $\delta$ is the half channel heigh, $P^*$ = P/($\rho_0 U^2$), $\rho_0$ is the inlet density, $\rho^*$ = $\rho/\rho_0$ = 1; $T^*$ = (T-$T_0$)/$\Delta T$, $T_0$ is the inlet temperature, $\Delta T$ is the temperature difference between inlet and outlet (for case 2 in Figure 3) or between the 2 walls (for case 1 in Figure 3). Note that Re = (DU$\rho_0$)/$\mu$, $Fr = U/\sqrt{gD}$, Pe = PrRe = ($\rho_0$UD$c_{p0}$)/k, Ri = $\beta\Delta T$/Fr$^2$, where Fr is the Froude number.

In order to characterize the mix convection phenomena, specifically, the relative strength between forced and natural convection, The Richardson number Ri defined quantitatively the force term in equation (5). Ri is defined by the ratio of Grashof number (7) to the square of Re number, as show in (8):

$$Gr = \frac{\rho^2 g \beta \Delta T D^3}{\mu^2} \tag{7}$$

$$Ri = \frac{Gr}{Re^2} = \frac{\beta \Delta T}{Fr^2} \tag{8}$$

For the sCO$_2$ case, NekRS is used for all simulations because using GPUs can lead to significant speed-ups (up to 40 times compared to Nek5000 using CPUs [21]) and The NS nondimensional form is deployed to take advantage of physical parameters, tolerances, etc. in prior DNS of sCO$_2$ by Nek5000 [22]. The Low-Mach approximation [23] has been applied to address the working fluid's physical properties changes in agreement with the temperature variation in the circular pipe. In NekRS, the low-Mach-number approximation is achieved by adding a source term to the mass continuity equation (1). The difference between low-Mach Navier-Stokes's equations and incompressible Navier-Stokes's equations is the source term in (1) which accounts for the change in density. The equations (4), (5), (6) now becomes:

$$\nabla \cdot u^* = -\frac{1}{\rho^*}\left(\frac{\partial \rho^*}{\partial t^*} + u^* \cdot \nabla \rho^*\right) \tag{9}$$

$$\rho^* \left(\frac{\partial u^*}{\partial t^*} + u^* \cdot \nabla u^*\right) = -\nabla P^* + \nabla \cdot \frac{1}{Re}(\nabla(u^*) + \nabla(u^*)^T) + Fr^{-2}\rho^* \tag{10}$$

$$\rho^* \left(\frac{\partial h^*}{\partial t^*} + u^* \cdot \nabla h^*\right) = \nabla \cdot \frac{k^*}{c_p^*} \frac{1}{Pe} \nabla h^* \tag{11}$$

Where $c_p^*$, $k^*$, $h^*$ are nondimensional heat capacity, conductivity, enthalpy and Froude number, respectively. $k^* = k/k_0$, $k_0$ is the inlet conductivity, $c_p^* = c_p/c_{p0}$, $c_{p0}$ is the inlet heat capacity, $h^* = (h-h_0)/\Delta h$, $h_0$ is the inlet enthalpy, $\Delta h$ is the enthalpy difference between inlet and outlet of the pipe, $Fr = U/\sqrt{gD}$, Re = $(DU\rho_0)/\mu$, Pe = $(\rho_0 U D c_{p0})/k_0$.

It's important to note that for sCO$_2$, some properties vary significantly near the pseudo-critical point, which can cause difficulties for numerical simulation if temperature is used as the primitive variable in the energy equation. Therefore, enthalpy is used instead of temperature in (11).

The added source term in mass continuity equation is calculated from the energy equation:

$$\nabla \cdot u^* = -\frac{1}{\rho^*}\left(\frac{\partial \rho^*}{\partial t^*} + u^* \cdot \nabla \rho^*\right) = \frac{1}{(\rho^*)^2}\frac{\partial \rho^*}{\partial h^*}\left(\nabla \cdot \frac{k^*}{c_p^*}\frac{1}{Pe}\nabla h^*\right) \tag{12}$$

From (12), $\partial \rho^*/\partial h^*$ must be calculated for the properties of sCO$_2$. Details of the implementation of the low-Mach-number formulation are available in [23]. The velocity and temperature are averaged in the azimuthal direction to reduce convergence time. For a given axial and radial position, for any variable, its Reynolds averaged value is defined as following:

$$\langle \alpha \rangle = \frac{1}{T}\int_t \left[\frac{1}{2\pi}\int_\varphi \alpha(x,r,\varphi)d\varphi\right]dt \tag{13}$$

## 2.2 Numerical setup

### 2.2.1 Simplified Downcomer (parallel plates)

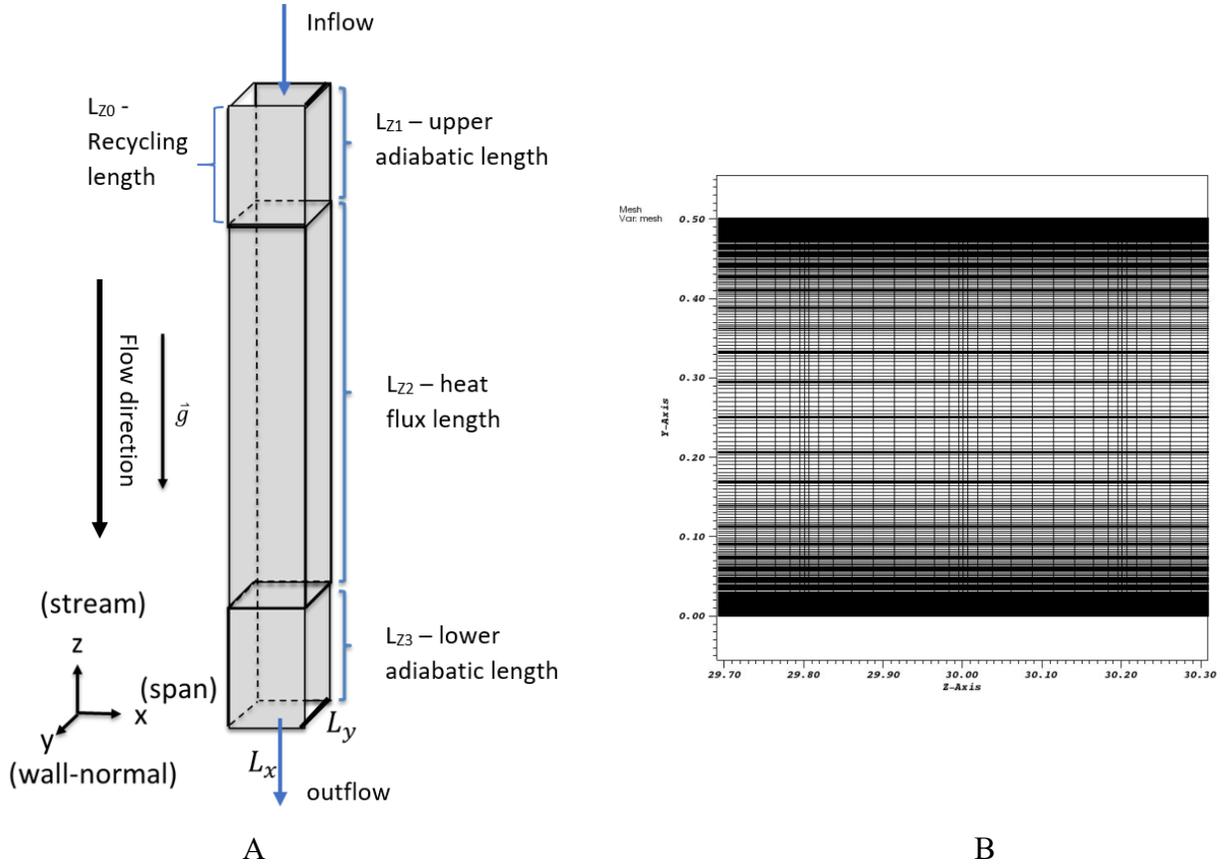

**Figure 2.** A – parallel plates computational box. B - The grid employed in the simulation from half of the channel's cross section for Re = 5000. 10$^{th}$ order.

The computational model for the parallel plates case is shown in Figure 2. The fluid domain is divided into 3 sections: upper adiabatic, heat flux and lower adiabatic. The ratio between them, as well as the spanwise $L_x$ and wall normal $L_y$ lengths are defined according to the interest to our industrial partner (Kairos Power).

$$L_x = \pi \, L_y \tag{14}$$

$$L_{Z1} = L_{Z3} = 10 L_y \tag{15}$$

$$L_{Z2} = 100 L_y \tag{16}$$

Based on (14), (15), (16), the streamwise length of the channel is $120 L_y$. To achieve a fully developed turbulent velocity profile before entering the heat flux section, a recycling method is applied with the length of $L_{Z0} = L_{Z1}$. As a result, a fully-developed inflow condition to the head flux section at desired Re are generated.

Gmesh [24] - an open-source 3D finite element mesh generator, has been used to generated the mesh. The resulting mesh has 250000 hexahedra for Re = 5000. At $7^{th}$ polynomial order, the computational domain obtained 128 million degrees of freedom. The maximum $y^+$ are lower than 0.1 along the walls. The discretization has satisfied standards for the DNS of channel flow, which includes $y^+ < 1$ near wall, 10 grid points below $y^+ = 10$, $\Delta x^+ < 4$ and $\Delta z^+ < 8$. We also investigated mesh convergence by comparing simulation results for the $8^{th}$, $9^{th}$ and $10^{th}$ polynomial order (as shown in Figure 2) and we observed no meaningful differences in generated turbulence data. The time step in the simulations is set to a maximum of $10^{-3}$ s (CFL<0.5) and the gravitational direction was chosen to be on the z axis, as shown in Figure 2.

We investigated two cases (case 1 and case 2) as shown in Figure 3 representing different cooling conditions at the wall boundary. The working fluid is FLiBe at 695 $^0$C which corresponds to Pr = 12 in nondimensional terms. Neuman boundary condition (BC) imposing constant heat flux are applied for two opposite vertical walls in +y and -y direction. The spanwise direction +x and -x are periodic. The Dirichlet BC imposing constant temperature are used at inlet.

The heat flux applied to each heated plate is also nondimensionalized, it is defined by (17):

$$f^* = f \cdot \left(\frac{1}{\rho_0 U c_p \Delta T}\right) \tag{17}$$

Where $f^*$ and $f$ are nondimensional and dimensional heat flux, respectively.

For case 1, the temperature difference $\Delta T$ between the walls is defined by (18)

$$\Delta T = \frac{f * D}{2k} \tag{18}$$

Hence, the nondimensional heat flux will becomes:

$$|f^*| = |f| \cdot \left(\frac{1}{\rho_0 U c_p \Delta T}\right) = \frac{2k}{\rho_0 U c_p D_h} = \frac{1}{RePr} = \frac{2}{Pe} \tag{19}$$

And the Ri from (8) will become:

$$Ri = \frac{g\beta |f| D^2}{2kU^2} \tag{20}$$

For case 2, the temperature difference $\Delta T$ between inlet and outlet, the nondimensional heat flux and Ri number are defined by (21), (22), (23) respectively.

$$\Delta T = \frac{|Q|}{\dot{m} c_p} = \frac{f * 2 L_{Z2}}{\rho_0 U c_p L_y} \tag{21}$$

Where Q is the heat energy, $\dot{m}$ is the mass flow rate.

$$|f^*| = |f| \cdot \left(\frac{1}{\rho_0 U c_p \Delta T}\right) = \frac{L_y}{2 L_{Z2}} \tag{22}$$

$$Ri = \frac{g\beta|f|D}{\rho_0 U^3 c_p} \tag{23}$$

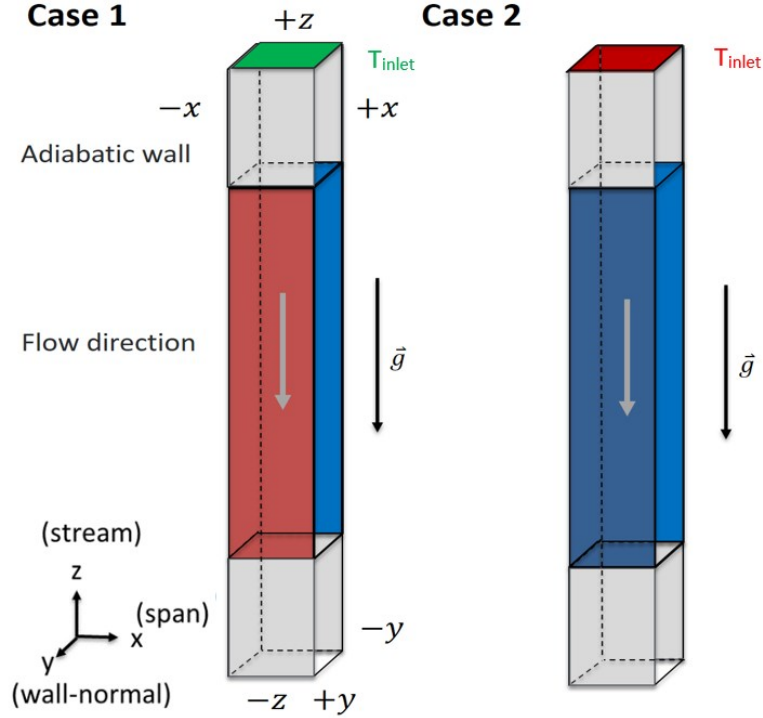

**Figure 3.** Boundary condition map for investigated cases in parallel plates.

The detail of boundary conditions (BCs) for case 1 and case 2 are summarized in table 1.

| BCs \ Geometry | Velocity BCs | | temperature BCs | |
|---|---|---|---|---|
| | Case 1 | Case 2 | Case 1 | Case 2 |
| +x and -x | Periodic | Periodic | Periodic | Periodic |
| +y, for $L_{z1}$ and $L_{z3}$ | No-slip | No-slip | Isolated | Isolated |
| +y, for $L_{Z2}$ | No-slip | No-slip | $f^* = 2/Pe$ | $f^* = -L_y/(2L_{Z2}) = -1/200$ |
| -y, for $L_{z1}$ and $L_{z3}$ | No-slip | No-slip | Isolated | Isolated |
| -y, for $L_{Z2}$ | No-slip | No-slip | $f^* = -2/Pe$ | $f = -L_y/(2L_{Z2}) = -1/200$ |
| +z | inflow | inflow | $T^* = 0$ | $T^* = 0$ |
| -z | Pressure outflow | Pressure outflow | Isolated | Isolated |

Table 1: the velocity and temperature BCs applied for case 1 and case 2

The initial conditions are the same for all cases,

$$T^* = 0 \tag{24}$$

$$u_x^* = u_y^* = u_z^* = 0 \tag{25}$$

The summary of simulations cases is reported in table 2. Case 2 at Re = 9120 is performed for validation purposed.

| Re | Pr | Cases | | | |
|---|---|---|---|---|---|
| | | Case1 | | Case2 | |
| | | Gr | Ri | Gr | Ri |
| 5000 | 12 | 0 | 0 | 0 | 0 |
| 5000 | 12 | $2*10^9$ | 80 | $10^7$ | 0.4 |
| 5000 | 12 | $10^{10}$ | 400 | $5*10^7$ | 2 |
| 9120 | 0.71 | - | - | 0 | 0 |

Table 2: the simulation Cases with their corresponding Re, Gr and Ri number. Ri = 0 means force convection cases.

### 2.2.2. Pipe in mixed convection with sCO$_2$

For the sCO$_2$ test case the geometry is a vertical straight heated circular tube. A 96,000 hexahedra mesh has been generated using Gmsh. At 7$^{th}$ polynomial order, the computational domain obtained ≈ 33 million degrees of freedom. The maximum y+ are lower than 0.1 along the walls. The mesh size has been confirmed to be satisfied Kolmogorov length scales, and the mesh near wall is well refined to have an accurate prediction of wall temperature, as shown in Figure 4. The time step is set at $5*10^{-6}$s and the gravitational direction was chosen to be on the z axis, and a recycling method is also applied with a length of L/D = 5, where L is the length of the tube, as shown in Figure 5.

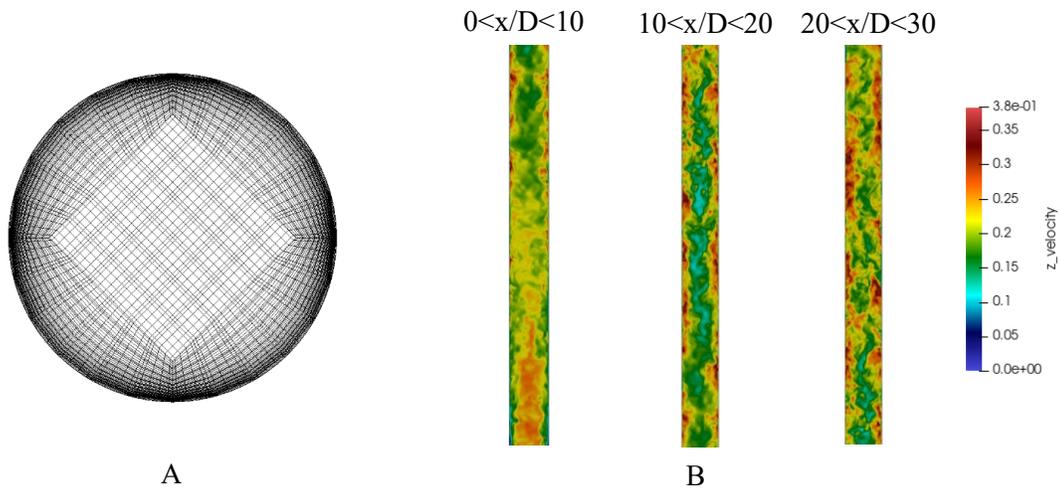

**Figure 4.** A – The Z-plane cross section of the tube's mesh for 7$^{th}$ Polynomial order, B – Z-plane instantaneous velocity field in Z direction of case 6 in table 3.

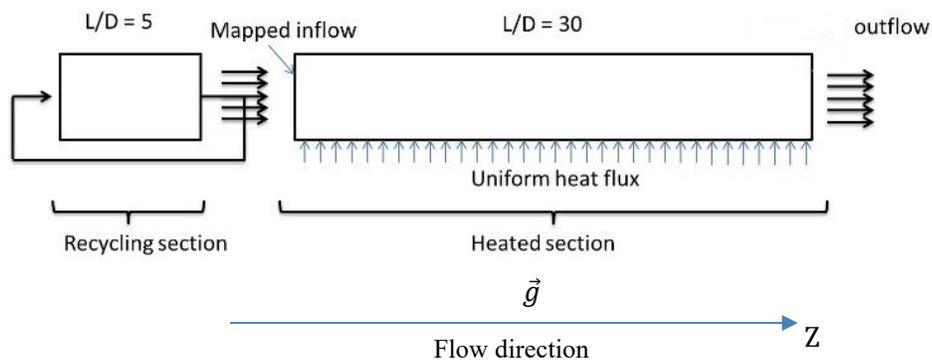

**Figure 5.** Tube's computational domain and boundary conditions

The nondimensional heat flux is defined in (26), which is similar to (17) but using enthalpy instead of temperature:

$$f^* = f \cdot \left(\frac{1}{\rho_0 U \Delta h}\right) \quad (26)$$

There are 6 cases corresponding to Bae's paper [11] that were performed for wall temperature evaluation, the details are in table 3.

| Case | Type/direction | Diameter (mm) | Wall heat flux (W/m$^2$) |
|------|----------------|---------------|--------------------------|
| 1 | Force | 1 | 61740 |
| 2 | Mix/downward | 1 | 61740 |
| 3 | Mix/upward | 1 | 61740 |
| 4 | Mix/downward | 2 | 30870 |
| 5 | Mix/upward | 2 | 30870 |
| 6 | Mix/upward | 2 | 61740 |

Table 3: Cases in Bae's paper with corresponding parameters

The inlet recycling velocity is 0.4454 (for cases at D=1mm) and 0.2227 (for cases at D=2mm) corresponding to Re ≈5300. In NekRS, case parameters in table 3 are nondimensionalized by sCO$_2$ reference properties at 8 MPa and 301.15 K. The solutions from NekRS are converted back to dimensional values, normalized by T$_o$ = 301.15K and then compared with the data in Bae's paper.

## 3. Results and Discussion

### 3.1. Validation

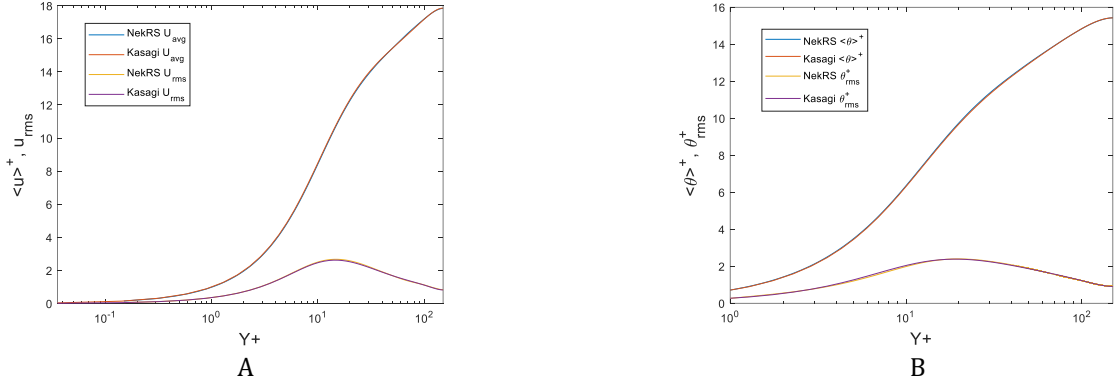

**Figure 6.** A – average and RMS streamwise velocity, B – average and RMS local temperature difference of NekRS and Kasagi et. al.

Currently, there is no applicable references for the relevant conditions of interest, at least for mixed convection cases (Ri ≠ 0) in table 2. A case 2 at Re = 9120, Pr = 0.71 has been performed to compare with Kasagi et.al [5,6] DNS data. The Re and Pr has been chosen to match with Kasagi et.al case. We note that in the work of Kasagi et. al., the hydraulic diameter is defined by D = 2δ, δ is the half channel heigh. In our work, D = 4δ which makes Re number increase by twice in comparison with Kasagi et. al. Turbulence statistic parameters including average streamwise velocity $<u>^+$, RMS streamwise velocity ($u^+_{rms}$), average local temperature difference $<\theta>^+$, RMS local temperature difference $\theta^+_{rms}$, as well as turbulence kinetic energy (TKE) budgets in streamwise direction are compared with previous DNS data and excellent agreements have been achieved as shown on Figure 6 and Figure 7. The agreement with Kasagi DNS data could only be obtained with certain runtime settings, which are then applied for all other cases in table 2.

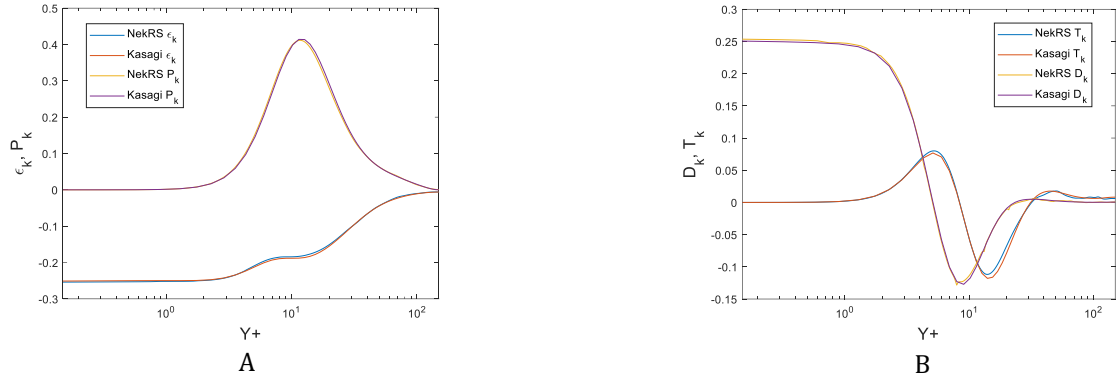

**Figure 7.** TKE budgets: A – dissipation term $\epsilon_k$ and Production term $P_k$, B – viscous diffusion term $D_k$ and turbulence diffusion term $T_k$ of NekRS and Kasagi et. al.

### 3.2. Pipe with $sCO_2$

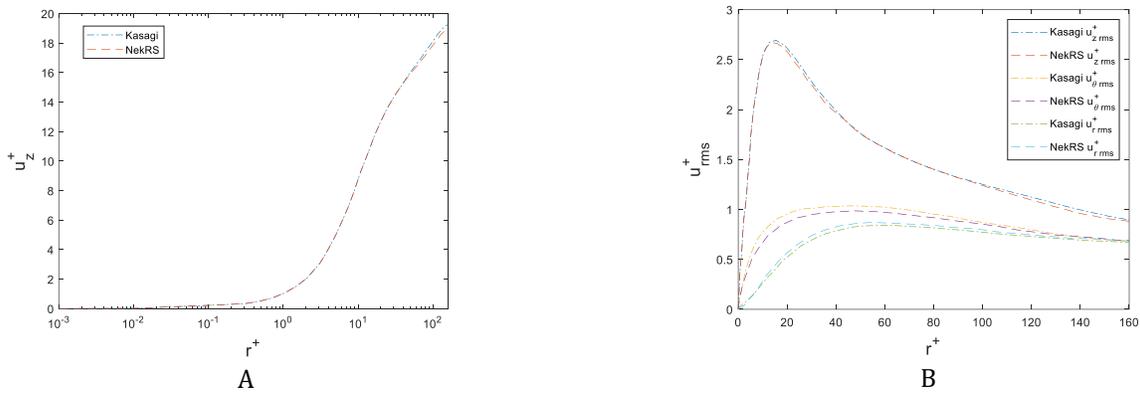

**Figure 8.** A – $sCO_2$ average streamwise velocity, B – $sCO_2$ rms velocity profile ($u_{z\,rms}^+$, $u_{\theta\,rms}^+$, $u_{r\,rms}^+$) of Kasagi et. al. and NekRS.

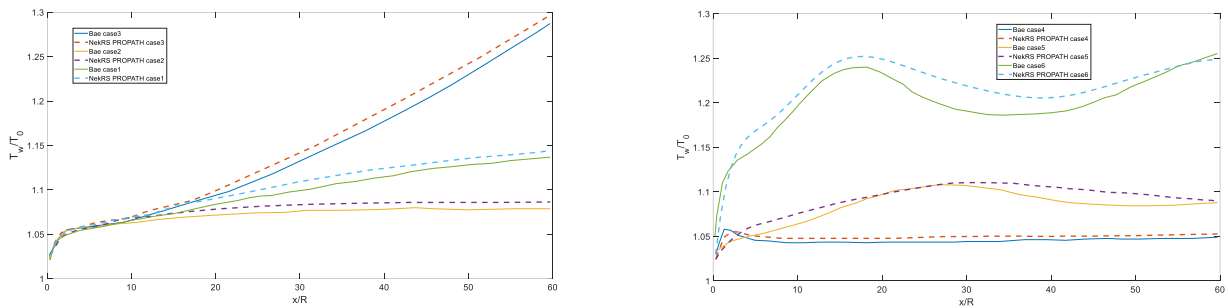

**Figure 9.** $sCO_2$ average wall temperature distributions for cases in table 3 of NekRS and Bae et. al, using PROPATH properties plugin.

For the $sCO_2$ case, the sCO2 properties changes in Low-Mach approximation are solved with the Program Package for Thermophysical Properties of fluids (PROPATH) [25] databases, which is the same as in Bae's work. PROPATH is a software library written by PROPATH team. The inlet velocity profile before entering the heated section has been compared to with data from Kasagi et. al. [5] and good agreement has also been achieved, as shown on Figure 8. For all cases in table 3, NekRS can reproduce the wall temperature of Bae DNS data with adequate agreement, as shown on Figure 9. Another $sCO_2$ data base (REFPROP) [26] has also been used to calculate the average wall temperature in table 3. The difference between REFPROP and PROPATH, as well as the simulation results comparison are available at [21].

### 3.3. Heated parallel plates at different Richardson and Reynolds numbers.

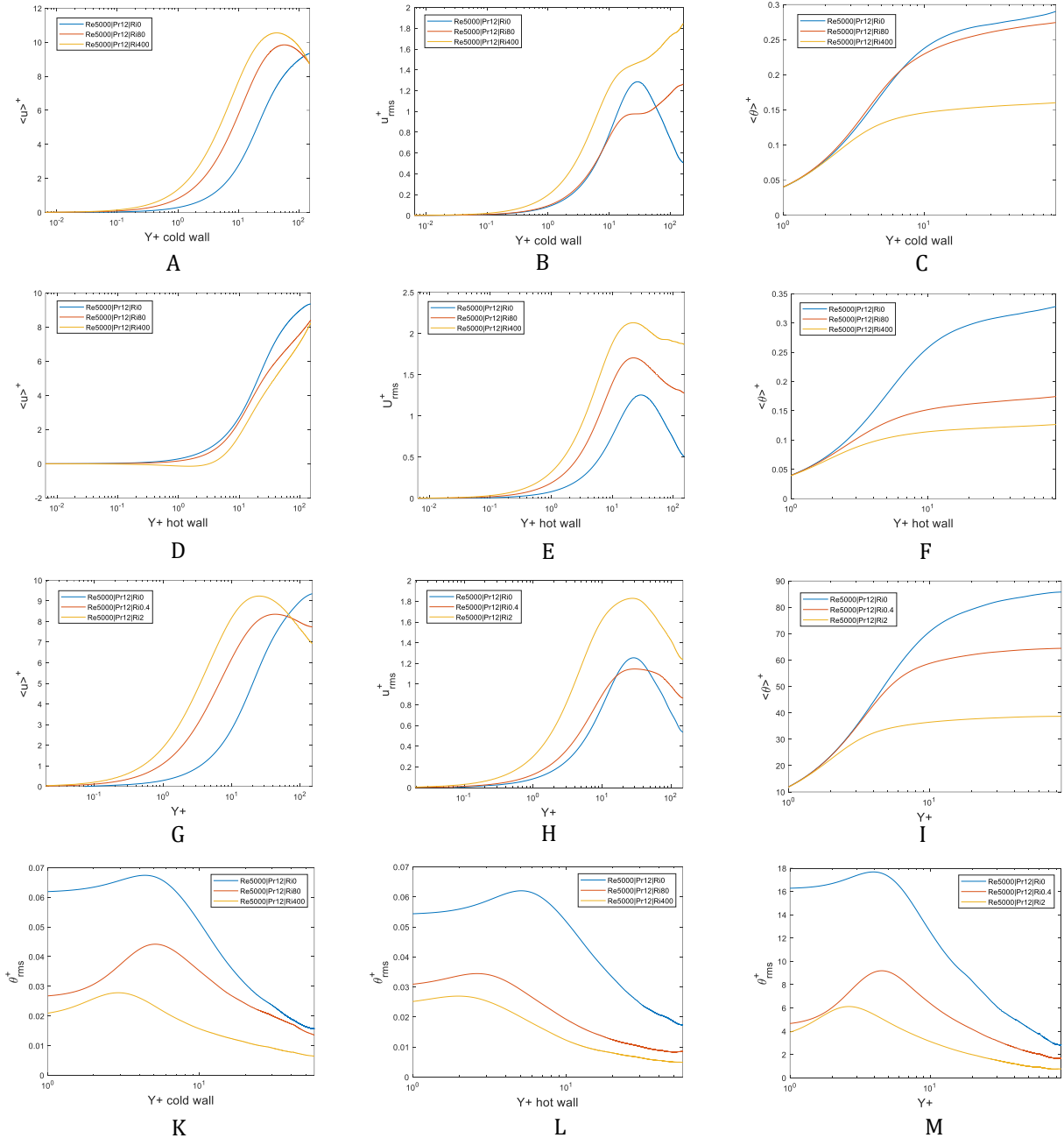

**Figure 10**. A, D – average streamwise velocity $\langle u \rangle^+$ for cases 1 (Ri = 0, 80, 400). G – $\langle u \rangle^+$ for cases 2 (Ri = 0, 0.4, 2.0). B, E – RMS streamwise velocity $u^+_{rms}$ for cases 1. H – $u^+_{rms}$ for cases 2. C, F – average local temperature difference $\langle \theta \rangle^+$ for cases 1. I – $\langle \theta \rangle^+$ for cases 2. K, L – RMS local temperature difference $\theta^+_{rms}$ for cases 1. M – $\theta^+_{rms}$ for cases 2.

In this section, we investigate the effect of buoyancy in mix convection cases at Re = 5000 for different Richardson (Ri) number for case 1 (Ri = 80, 400) and case 2 (Ri = 0.4, 2.0) as referenced in Table 2. A comparison campaign of turbulence statistics has been conducted between forced convection cases (Ri = 0) and mixed convection cases (Ri ≠ 0) for both case 1 and case 2 (Figures 10 and 11). We also collected time series of streamwise velocity of points near wall and at the center of the channel for cases 1 (Ri = 0 and Ri = 400) and cases 2 (Ri = 0 and Ri = 2.0), the results are shown in Figure 12. All the turbulence data has been collected on a z-plane cross section at $z = L_{Z1} + 0.5L_{Z2} = 60L_y$, which is at the middle of the streamwise length of the channel. Noted that for forced convection cases, the velocity turbulence statistics are the same for case 1 and case 2. For cases 1, turbulence statistics need to be obtained for both hot and cold wall due to the asymmetric temperature BCs, which leads to opposite signs of buoyancy forces at hot wall and cold wall sides.

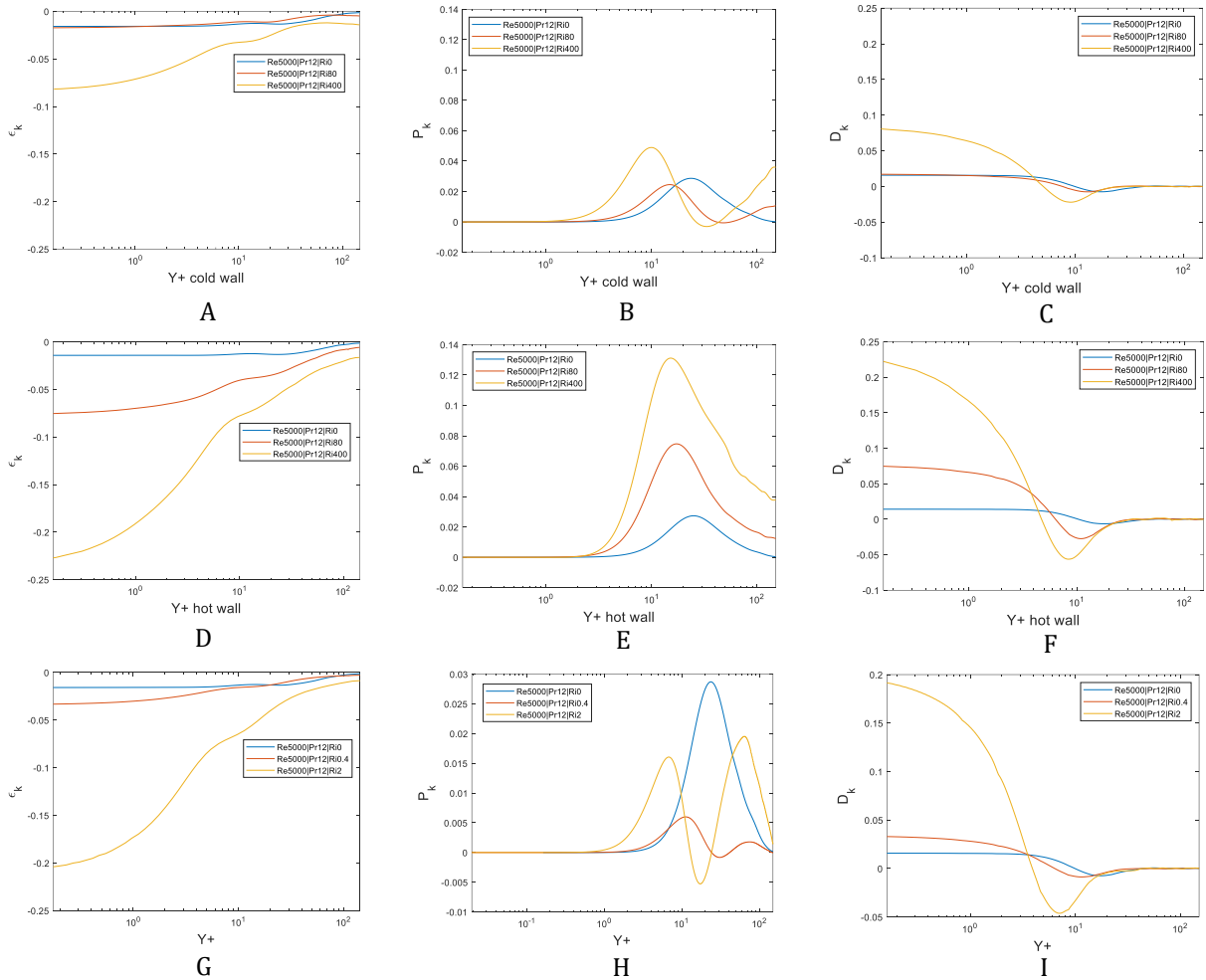

**Figure 11.** A, D – TKE dissipation $\varepsilon_k$ for case 1. G – $\varepsilon_k$ for case 2. B, E – TKE Production $P_k$ for case1. H – $P_k$ for case 2. C, F – viscous diffusion $D_k$ for case 1. I – $D_k$ for case 2.

In general, the average velocity and temperature profile, as well as the TKE budgets dramatically change as the Ri number increases and the flow transitions from forced to natural convection. The behavior of turbulence profiles of case 2 is similar to that of the cold wall side in case 1. For case 1, the RMS of velocity and temperature $u^+_{rms}$, $\theta^+_{rms}$ of the cold wall side is lower than that of the hot wall side, as shown on Figure 10 (B, E, H, K, L, M). In fact, on the cold wall side, the buoyancy force accelerates the flow. Whereas, on the hot wall side, it pushes again the flow, leading to increased dissipation, high TKE levels and chaotic flow structure. The effect of inertial forces is particularly evident in the production term $P_k$. Case 2 shows a shift from single peak (forced convection) to double peak (mix convection) as shown in Figure 11 (H), which indicate that massive $P_k$ happens not only close to the wall but also near the center of the channel for mix convection. But sufficiently high Ri number, both walls in case 2 and the cold wall in case 1 have regions with negative production $P_k$. This is typical of conditions with strong flow acceleration due to the buoyancy force (Figure 11 B, H). The bigger Ri is, the stronger the negative $P_k$ and the closer to the wall it is. On the other hand, the production of the hot wall for case 1 is significantly higher than the cold wall, which can be explained by the opposite direction of the buoyancy force in this region (Figure 11 E). This also can explain the differences between the hot and cold wall of case 1 for other TKE budgets terms, in particular, the amplitude of dissipation $ɜ_k$, viscous diffusion $D_k$ are about 4 times larger on the hot wall side in comparison with the cold wall side, as shown in Figure 11 (A, D, G, C, F, I).

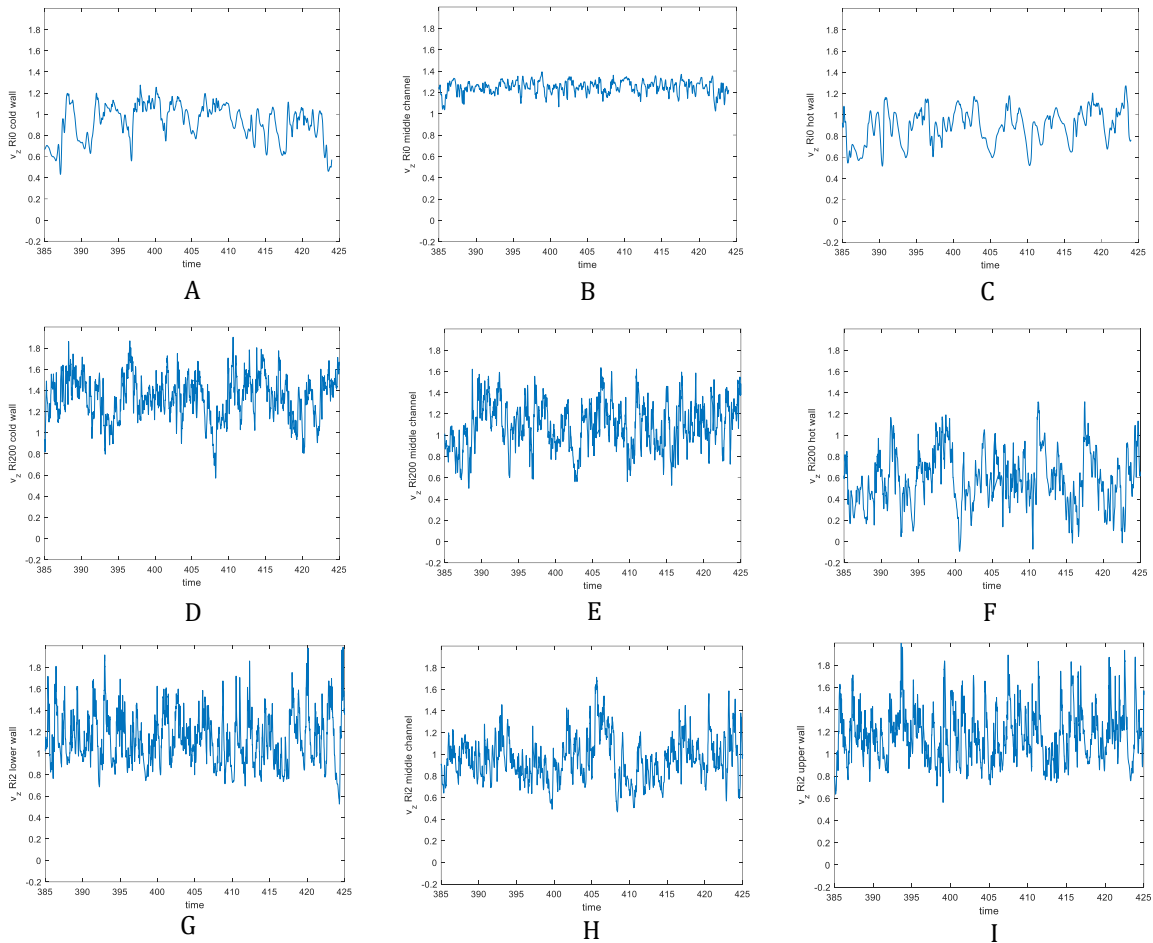

**Figure 12.** A, B, C – time series of streamwise velocity $v_z(t)$ for forced convection cases. D, E, F – $v_z(t)$ for case 1 at Ri = 400. G, H, I – $v_z(t)$ for case 2 at Ri = 2.0.

Let us now consider the time signals of streamwise velocity for case 1 (Ri = 0 and Ri = 400) and case 2 (Ri = 0 and Ri = 2.0). Figure 12 shows that the oscillation frequency and amplitude of $v_z$ increase drastically from forced (A, B, C) to mix convection cases (D, E, F, G, H, I). We also observe the presence of a low frequency content as the Ri number increases (case 1 at Ri = 400 and case 2 at Ri = 2.0). The center point of case 1 shows a stronger oscillation amplitude than that of case 2. The large oscillation amplitude of $v_z$ combined with the higher low frequency content for the points near wall of cases 2, as well as for the point near hot wall of case 1 signals the presence of large-scale vortices. This is consistent with what observed visually near the walls, as shown on Figure 1. These larges vortices are absent for the forced convection case. More detailed statistical analysis (e.g., POD), as well as PSDs of the time series will be conducted for an in-depth analysis of these flow structures.

## 4. Conclusion

We investigated forced and mixed convection heat transfer for a variety of conditions demonstrating the advanced capabilities of the spectral element codes Nek5000 and NekRS with Direct Numerical Simulation (DNS). We focus on two canonical test cases.

The primary case includes the simulation of high Pr fluid (FLiBe) between heated parallel plates, presentative of the downcomer of advanced reactor designs. At first, we have performed a DNS of a low Pr fluid case then compared with the Kasagi DNS data and good agreement has been achieved. The runtime settings have then been applied for cases at Re = 5000 with a wide range of Ri number from 0 to 400 and turbulence statistics data for velocity and temperature has been collected. The simulation results shows that there is a dramatic transition in average velocity and temperature profiles, as well as the RMS and TKE budgets as the Ri number increases. The behavior of turbulence profiles for cases with symmetric cooling BCs is similar to cases with asymmetric cooling on the cold wall side. The TKE budgets terms, show amplitudes at the hot wall side that are about 4 times larger than the cold wall side. Furthermore, we observe negative production region near the cold walls, where the flow is accelerated by the buoyancy force. On the hot wall side, the buoyancy force is directed against the flow, which leads to higher levels of energy production. The streamwise velocity time signal at points near the hot wall at high Ri number show evidence of low frequency content associated with large scale structures.

We also performed a series of DNS simulation for $sCO_2$ in straight heated tube involving mixed convection using NekRS. $sCO_2$ properties, which change rapidly with enthalpy, have been obtained using the PROPATH database and implemented using a low-Mach approximation. The wall temperature calculation results have been compared with Bae's data and good agreement has been archived for both 1mm and 2mm diameters tube cases. The simulations results confirmed that NekRS has capabilities to simulate mixed convection even when considering the full dependency of properties as a function of temperature or enthalpy with good accuracy and efficiency.